\documentclass[a4paper,conference]{IEEEtran}
\ifCLASSINFOpdf
\else
\fi
\ifCLASSOPTIONcompsoc
 \usepackage[caption=false,font=normalsize,labelfont=sf,textfont=sf]{subfig}
\else
 \usepackage[caption=false,font=footnotesize]{subfig}
\fi
\usepackage{url}

\usepackage[english]{babel}

\usepackage{xcolor}

\usepackage{soul}

\usepackage[textsize=tiny,colorinlistoftodos]{todonotes}

\usepackage[range-phrase={\,--\,}, range-units=single, per-mode=symbol, detect-all]{siunitx}

\makeatletter
  \@addtoreset{paragraph}{section}
\makeatother

\usepackage[acronym,shortcuts]{glossaries}
\newacronym{vr}{VR}{Virtual Reality}
\newacronym{xr}{XR}{eXtended Reality}
\newacronym{qoe}{QoE}{Quality of Experience}
\newacronym{qos}{QoS}{Quality of Service}
\newacronym{uav}{UAV}{Unmanned Aerial Vehicle}
\newacronym{faas}{FaaS}{Function-as-a-Service}
\newacronym{pubsub}{Pub/Sub}{Publish/Subscribe}
\newacronym{h3d}{H3D}{holographic 3D}
\newacronym{ai}{AI}{Artificial Intelligence}
\newacronym{nbmp}{NBMP}{network-based media processing}
\newacronym{kpi}{KPI}{key performance indicator}
\newacronym{hc}{\emph{HoloCom}}{\emph{Holographic Communication}}
\newacronym{ml}{ML}{Machine Learning}
\newacronym{rem}{REM}{radio environment map}
\newacronym{drl}{DRL}{deep reinforcement learning}
\newacronym{sdr}{SDR}{software-defined radio}
\newacronym[firstplural=radio access technologies]{rat}{RAT}{radio access technology}
\newacronym{ran}{RAN}{radio access network}
\newacronym{tcas}{TCAS}{traffic alert and collision avoidance system}
\newacronym{ga}{GA}{general aviation}
\newacronym{mcs}{MCS}{Modulation and Coding Scheme}
\newacronym{fec}{FEC}{forward error correction}
\newacronym{api}{API}{application programming interface}
\newacronym{lstm}{LSTM}{long short term memory}
\newacronym{cnn}{CNN}{Convolutional Neural Network}
\newacronym{cam}{CAM}{Class Activation Map}

\usepackage[
        bibstyle=ieee,
        citestyle=numeric-comp,
        maxcitenames=6,
        doi=false,
        eprint=false
    ]{biblatex}
        
\AtEveryBibitem{
  \clearfield{urlyear}%
  \clearlist{language}%
 }

\usepackage{balance}
\usepackage[capitalize]{cleveref}
\usepackage[all]{nowidow}
\usepackage{url}

\graphicspath{{figures/}}

\begin{document}
%

\title{6G NeXt - Split Computing for Smart Drones}


%
\author{
    \IEEEauthorblockN{
        Sergiy Melnyk\IEEEauthorrefmark{1},
        Qiuheng Zhou\IEEEauthorrefmark{1},
        Hans D. Schotten\IEEEauthorrefmark{1}\IEEEauthorrefmark{2},
        Tobias Pfandzelter\IEEEauthorrefmark{3},\\
        David Bermbach\IEEEauthorrefmark{3},
        Robert Vilter\IEEEauthorrefmark{4},
        Nick Stuckert\IEEEauthorrefmark{4},
    }
    \IEEEauthorblockA{
        \IEEEauthorrefmark{1}
        Intelligent Networks,
        German Research Center for Artificial Intelligence (DFKI), Kaiserslautern, Germany \\
        Email: \{sergiy.melnyk, qiuheng.zhou, hans\_dieter.schotten\}@dfki.de
    }
    \IEEEauthorblockA{
        \IEEEauthorrefmark{2}
        Institute for Wireless Communication and Navigation,
        University of Kaiserslautern (RPTU), Kaiserslautern, Germany\\
        Email: schotten@rptu.de
    }
    \IEEEauthorblockA{
        \IEEEauthorrefmark{3}
        Scalable Software Systems Research Group, Technische Universität Berlin, Berlin, Germany \\
        Email: \{tp, db\}@3s.tu-berlin.de
    }
    \IEEEauthorblockA{
        \IEEEauthorrefmark{4}
        Aviation Engineering Group, Technical University of Applied Sciences Wildau, Wildau, Germany \\
        Email: \{vilter, stuckert\}@th-wildau.de
    }    

}

\maketitle
\vspace{-3mm}  

\begin{abstract}
  With the rising amount of \acs{uav} missions for logistics, agriculture, and other domains, control and anti-collision applications will gain importance. In this paper, we propose a \acs{uav} coordination system following a split computing paradigm. Furthermore, a proof-of-concept setup was built, showing promising results for future utilisation of the proposed system.
\end{abstract}


\section{Introduction}

The utilisation of \glspl{uav} is propagating to new fields of operation such as logistics, agriculture, and natural disaster control. In Europe, the number of commercial \glspl{uav} is expected to reach 400\,000 by 2050~\cite{undertaking2017european}. This high amount of autonomous traffic will lead to an increasing risk of mid-air collisions, especially in dense air spaces such as airports, where \glspl{uav} are taking off and landing.

Research project \textit{6G NeXt}\footnote{\url{https://www.6gnext.de/}} addresses this issue~\cite{melnyk2023towards,melnyk2023joint}. Through a centralized anti-collision system that monitors the air traffic in a certain area, dangerous situations should be predicted and avoided. The high mobility and large coverage area of drones as well as latency constraints increase the overall complexity of the required system. Whereas fifth generation mobile networks provide a good starting point for the research activity, the requirements of the anti-collision application brings 5G to the limits, requiring further research on the next generation of mobile networks~\cite{eu5Gvision,itunetwork2030}.


\begin{figure}
    \centering
    \includegraphics[width=\linewidth]{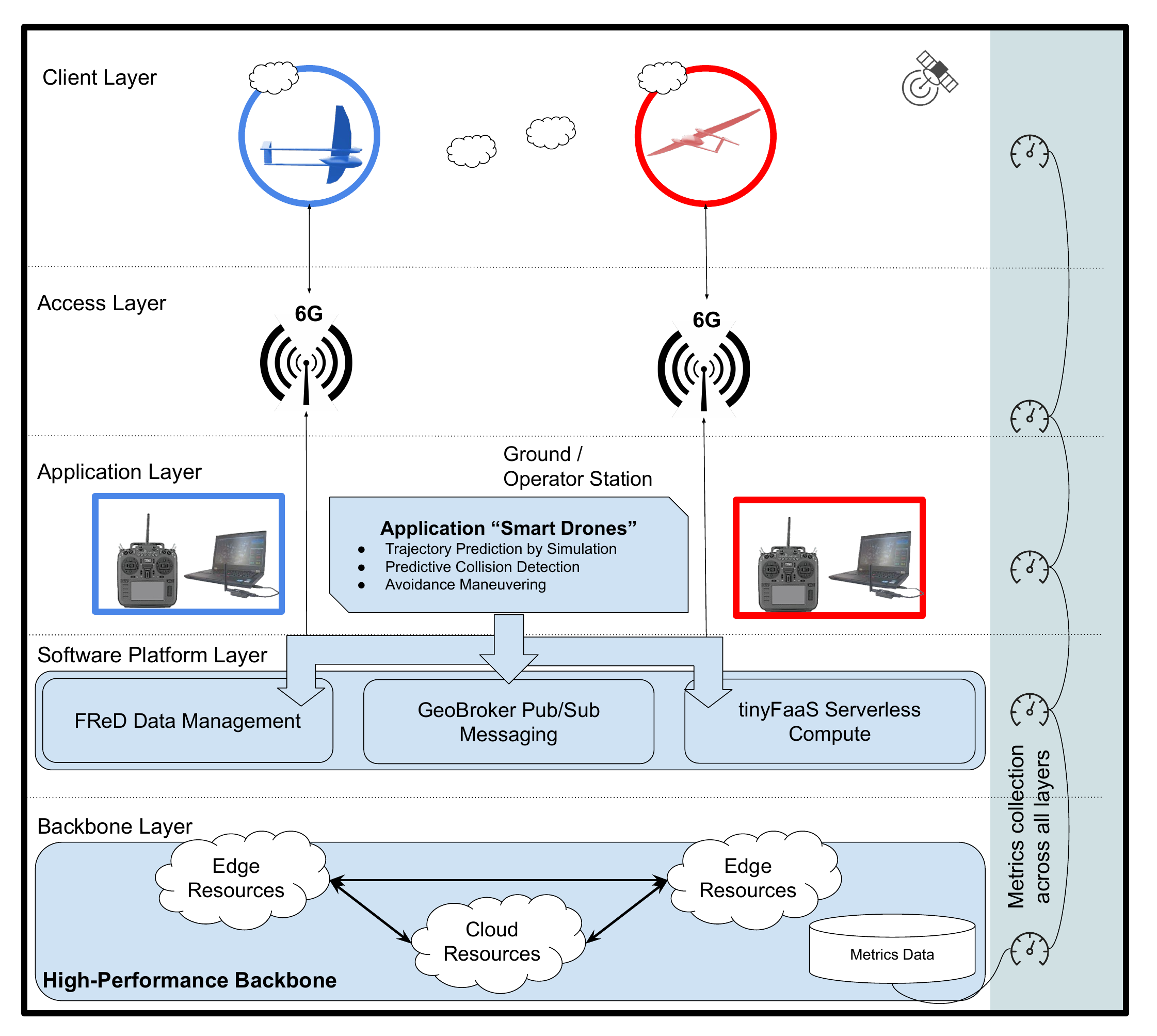}
    \caption{Architecture of a smart drones anti-collision system \cite{melnyk2023joint}}
    \label{fig:architecture}
\end{figure}


\begin{refsection}
    \let\thefootnote\relax\footnote{This is a preprint of a work accepted but not yet published at the 2025 European Conference on Networks and Communications (EuCNC ’25). Please cite as: \fullcite{melnyk20256g}.}
\end{refsection}%

\section{Smart Drones Anti-Collision System}

\begin{figure*}[t]
    \centering
    \subfloat[Flight path and airspeed]{%
        \includegraphics[width=.278\textwidth]{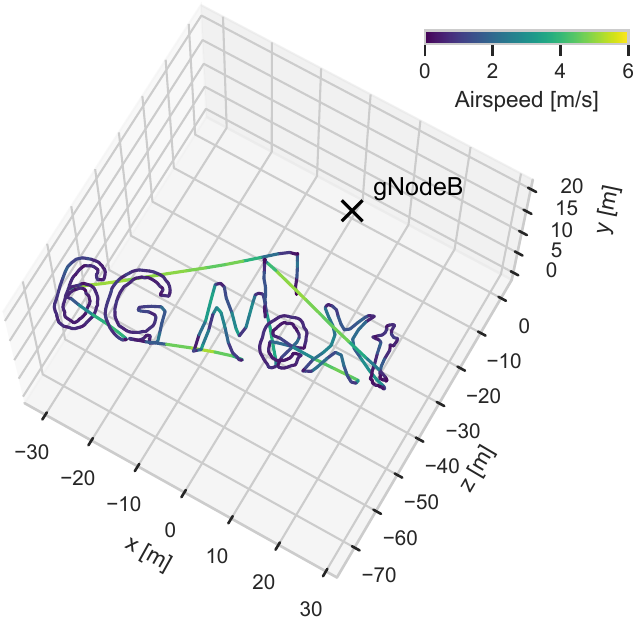}
        \label{fig:meas_path}
    }\hfill%
    \subfloat[Network latency measurement (\texttt{ping})]{%
        \includegraphics[width=.278\textwidth]{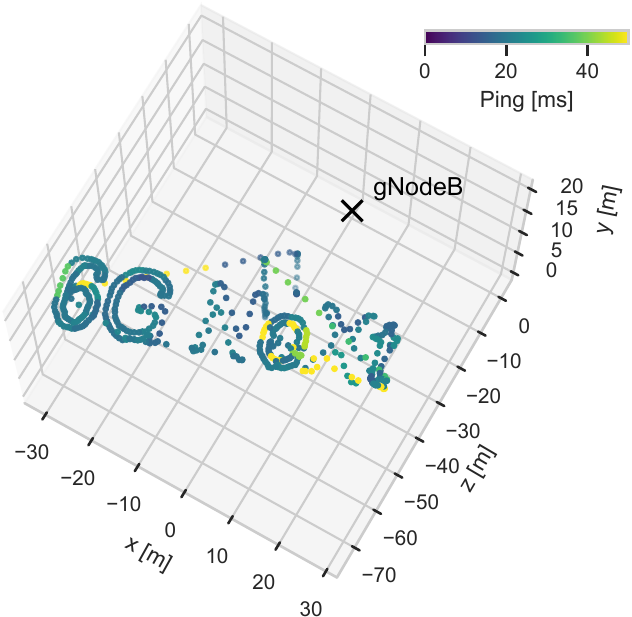}
        \label{fig:meas_ping}
    }\hfill%
    \subfloat[Network throughput (\texttt{iperf} receive)]{%
       \includegraphics[width=.278\textwidth]{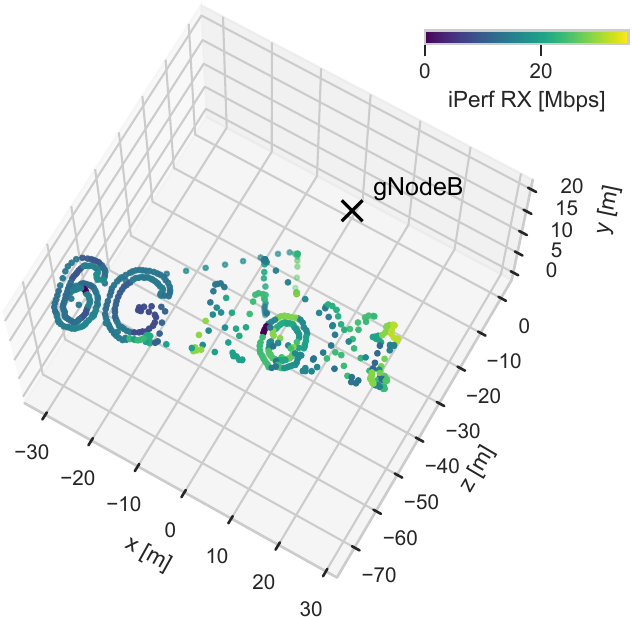}
        \label{fig:meas_iperf}
    }%
    \caption{Proof-of-concept measurement results}
    \label{fig:meas}
\end{figure*}

In our paper, we propose an anti-collision system for \glspl{uav}, which relies on predicting the flight paths of all vehicles in a certain area. On the one hand, the prediction algorithm is greedy for processing power. Since high processing power means also heavy equipment, it is advantageous for a \gls{uav} to offload the prediction computations to a ground-based server in order to reduce the take-off weight. On the other hand, the \glspl{uav} need to coordinate their flight maneuvers with each other. Thus, a centralized processing entity would also increase the level of coordination between the \glspl{uav}. For these reasons, we propose to split the control application into client and server part, as illustrated in \cref{fig:architecture}:

    \begin{itemize}
        \item \textbf{\gls{uav} Controller}: \gls{uav}-based flight controller, which requests the mission data from the Ground Station and reports the telemetry data (position, elevation, velocity, direction) back to it.
                
        \item \textbf{Ground Station}: ground-based simulation engine, which captures the flight data from \glspl{uav}, performs the anti-collision-prediction and commands the involved \glspl{uav} in case of harm probability.
    \end{itemize}

The connection between \glspl{uav} and the ground station is provided by a 6G mobile network, which should be able to provide the low-latency, stable, and reliable link for each involved \gls{uav}. Note that the large flight range of \glspl{uav} may require several base stations for continuous coverage, while their high speed makes the management of network traffic between \glspl{uav} and the ground complex. To mitigate this issue, we propose to utilize a distributed serverless edge-cloud infrastructure~\cite{pfandzelter2024towards}. Utilizing an additional geo-aware abstraction layer, the procedure of messaging as well as computational distribution can be mitigated.

\section{Proof of Concept}

According to the architecture proposed in the previous section, we set up all required systems to perform multiple proof-of-concept flight tests at Rangsdorf model air field in Germany. Our \gls{uav} is a custom-designed quadcopter equipped with a \emph{Cube Orange/+} flight controller and a \emph{Raspberry Pi 4B} with 5G/6G connectivity through a \emph{Quectel RM500} modem. For 5G/6G network connectivity, an OAI-based gNodeB setup for campus networks with an \emph{NI Ettus USRP X410} was used. Our ground station software ran on a local edge server.

During the flight, the \gls{uav} mission was entirely controlled by the ground station over the 5G/6G network. Besides mission waypoints, the \gls{uav} was instructed to perform mobile network latency (\texttt{ping}) and throughput (\texttt{iperf} downlink and uplink) tests over the 5G/6G link. Results and telemetry parameters, such as location and airspeed, were collected, as we show in \cref{fig:meas}.

The degradation of throughput and latency correlates with the distance between the base station and \gls{uav} as well as its airspeed. However, there exist some points where the link degradation cannot be explained with an increased distance. These spots may get critical, especially if at this position the drone would be commanded by a ground station to perform an anti-collision maneuver. Understanding these effects is paramount to mitigate dangerous situations.

\section{Conclusion \& Future Work}
\label{sec:conclusion}

In our paper, we described the idea of the integration of the future-oriented 6G network and advanced compute infrastructure to support a smart drone anti-collision system. We provide a brief overview of the required software and hardware components and how they interact with each other to provide the best performance to the overall system. Besides, we performed a proof-of-concept flight, during which a drone was successfully controlled solely via a stable 5G/6G link while performing network tests and reporting telemetry data.

The tests showed promising results on the quality of the wireless connection between the \gls{uav} controller and the ground station. Nevertheless, our measurements showed some spots with weak link condition, which should be further mitigated. Using channel state prediction, such spots might be indicated and the reliability of the link might be increased by proactively switching to a more robust link configuration scheme. This is a part of the future work.

Further, we plan to step beyond the proof of concept. Setting up a fully functional anti-collision system would provide the possibility to verify the requirements on latency and reliability of the communication link. Here, the anti-collision algorithms will be further adjusted to the link parameters to increase robustness to the link degradation, especially with higher speeds and a higher number of \glspl{uav} sharing the airspace.

\section*{Acknowledgements}


Funded by the Bundesministerium für Bildung und Forschung (BMBF, German Federal Ministry of Education and Research) -- 16KISK176, 16KISK177, 16KISK183, 16KISK186.

\balance

\printbibliography

@INPROCEEDINGS{pfandzelter2024towards,
  author={Pfandzelter, Tobias and others},
  booktitle={Proc. IC2E}, 
  publisher = "IEEE",
  title={Towards Anti-Collision Coordination for UAVs with Serverless Edge Computing}, 
  year={2024},
  volume={},
  number={},
  pages={247-248},
  doi={10.1109/IC2E61754.2024.00035}
}

@inproceedings{melnyk2023joint,
    author = "Melnyk, Sergiy and others",
    title = "6G NeXt --- Joint Communication and Compute Mobile Network: Use Cases and Architecture",
    booktitle = "Proc. KommA",
    pages = "62--71",
    month = nov,
    year = 2023
}

@inproceedings{melnyk2023towards,
    author = "Melnyk, Sergiy and others",
    title = "6G NeXt --- Towards 6G Split Computing Network Applications: Use Cases and Architecture",
    booktitle = "Proc. MKT",
    pages = "126--131",
    month = may,
    year = 2023
}

@article{eu5Gvision,
	title = {European Vision for the 6G
Network Ecosystem},
author = "{The 5G Infrastructure Association}",
	year = {2021},
        doi = {10.5281/zenodo.5007671}
}

@techreport{itunetwork2030,
    title={Driving forces and vision towards Network 2030},
    institution = "{ITU-T -- Telecommunication Standardization Sector of ITU}",
    year = 2020,
    number = {NET2030-DF},
}

@techreport{undertaking2017european,
    author = "{SESAR}",
    title = "European drones outlook study: Unlocking the value for Europe",
    institution = "European Union",
    number = "MG-01-17-281-EN-C",
    booktitle = MISSING,
    month = MISSING,
    year = 2017
}

@inproceedings{melnyk20256g,
    author = {Melnyk, Sergiy and Zhou, Qiuheng and Schotten, Hans D. Schotten and Pfandzelter, Tobias and Bermbach, David and Vilter, Robert and Stuckert, Nick},
    title = {6G NeXt - Split Computing for Smart Drones},
    booktitle = {2025 European Conference on Networks and Communications  (EuCNC\,'25)},
    month = jun,
    year = 2025,
    publisher = {IEEE},
    venue = {Pozna\'n, Poland},
    numpages = 2,
}



\end{document}